\documentclass[pra,twocolumn,showpacs,byrevtex]{revtex4}

\usepackage{graphicx}
\usepackage{dcolumn}
\usepackage{amsmath}
\usepackage{epsfig}
\usepackage{bm}
\usepackage{amssymb}
\usepackage{sidecap}

\usepackage[dvips]{hyperref}
\hypersetup{colorlinks=true,linkcolor=blue,citecolor=blue,urlcolor=black}

\begin{document}

\title{Hybrid photonic entanglement: Realization, characterization and applications}

\date{\today}
\author{Leonardo Neves}\email{lneves@udec.cl} 
\author{Gustavo Lima} 
\author{Aldo Delgado}
\author{Carlos Saavedra}

\affiliation{Center for Quantum Optics and Quantum Information, Departamento de F\'{\i}sica, Universidad de Concepci\'on, Casilla 160-C, Concepci\'on, Chile.}



\begin{abstract}  

We show that the quantum disentanglement eraser implemented on a two-photon system  from parametric down-conversion is a general method to create hybrid photonic entanglement, namely the entanglement between different degrees of freedom of the photon pair. To demonstrate this, we generate and characterize a source with tunable degree of hybrid entanglement between two qubits, one encoded in the transverse momentum and position of a photon, and the other in the polarization of its partner. In addition, we show that a simple extension of our setup enables the generation of two-photon qubit-qudit hybrid entangled states. Finally, we discuss the advantages that this type of entanglement can bring for an optical quantum network. 

\end{abstract}

\maketitle

\section{Introduction}
	Photonic entanglement based on spontaneous parametric down-conversion (SPDC) has been extensively used as a resource in quantum communication (QC) with demonstrations of quantum teleportation \cite{Bouwmeester97}, cryptography \cite{Jennewein00} and Bell inequality violation \cite{Weihs98}. The importance of SPDC can be attested by its ability of preparing the most common types of entanglement and flexibility of using a variety of photonic degrees of freedom (DOFs), both discrete and continuous. Restricting to discrete spaces, sources for entangled qubits and qudits (two- and $D$-dimensional systems) encoded in polarization \cite{Kwiat99,Lamas01}, time-bin \cite{Thew02,Riedmatten04} and spatial \cite{Neves07,Neves05,Rossi09}  DOFs have been demonstrated, as well as sources of multiqubit entanglement \cite{Wieczorek08} and hyperentanglement, which is the simultaneous entanglement in more than one DOF \cite{Barreiro05}.

	In spite of these several branches of development, the concept of hybrid photonic entanglement---the entanglement between different DOFs of the photon pair---has not received, to date, significant amount of attention. In 1991, a theoretical proposal for a Bell test \cite{Zukowski91} has addressed this subject. There, a pair of polarization-entangled photons were converted into a hybrid entangled state (HES) between the polarization of a photon and the path followed by the other, which was defined by a polarizing beam splitter. Recently, this proposal was experimentally realized \cite{Ma09}. Also recently, we have employed HESs to implement a quantum eraser and study this phenomenon in the context of optimal quantum state discrimination \cite{Neves09}. In our case, however, the hybrid entanglement was defined between the polarization of a photon and the transverse spatial modes  of its partner, created by a double slit. Although both works used HESs for studying specific fundamental topics in quantum mechanics, none of them have addressed the general aspects of how to build up a HESs source, and more importantly, if they are useful for QC tasks. In this article, we address these both aspects and show that: (i) a general method to create HESs can be completely described in terms of the Garisto and Hardy \emph{disentanglement eraser}  \cite{Garisto99} implemented on a two-photon system from SPDC, and (ii) HESs may have important applications, as for instance, the engineering of qubit-qudit entangled states and the faithful transmission of quantum information through an optical quantum network comprised of free-space and optical fiber channels.

	A disentanglement eraser is a class of quantum erasers which restore entanglement rather than just interference. It consists of at least three quantum systems, where two of them are entangled. After a controlled-\textsc{not} (\textsc{cnot}) gate, the third system becomes entangled with the others, and the initial entanglement between the first two systems is lost or ``diluted'' into the full three-system state. It can be restored, however, by erasing the ``which-state'' information provided by the third system. This is achieved either by undoing the \textsc{cnot} operation (reversible eraser) or by a suitable projection of this subsystem (irreversible eraser) \cite{Garisto99}. 

	We show that when either eraser is applied on two \emph{arbitrary} two-dimensional DOF of a photon pair from SPDC, we get a two-qubit HESs source. The differences that arise in the source created with the reversible or irreversible eraser are discussed. To demonstrate the method, we implement the irreversible eraser on  polarization and (discretized) transverse momentum and position of down-converted photon pairs. In this way, we generate a source with tunable degree of hybrid entanglement which is fully characterized through quantum state tomography. We then demonstrate the usefulness of hybrid entanglement, and our setup in particular, by showing that it enables the generation of qubit-qudit entanglement in a fashion much more simple, flexible (with respect to the qudit dimension), and less costly than a previous scheme. Finally, we discuss the advantages that this type of entanglement can bring for an optical quantum network. 

	The remainder of this paper is organized as follows. In Sec.~\ref{sec_theory}, we describe the general theory behind the HES generation and our particular implementation. In Sec.~\ref{sec_exp}, we present the experiment results. In Sec.~\ref{sec_app} we discuss possible applications of hybrid photonic entanglement. A summary of our work is given in Sec.~\ref{sec_conc}.

\section{Theory} \label{sec_theory}

\subsection{General description} \label{sec_theory_a}
Figures~\ref{fig:setup}(a) and \ref{fig:setup}(b) illustrate the schemes of the irreversible and reversible disentanglement erasers, respectively, which create HESs from a SPDC source. A downconversion crystal generates photon pairs, signal ($s$), and idler ($i$), which are entangled in an arbitrary two-dimensional degree of freedom (DOF~$\mathbf{1}$). We call this the DOF control. A second DOF ($\mathbf{2}$) is separable from $\mathbf{1}$ and we call it the DOF target, which may or may not be entangled. To simplify the description here, we assume that it is not entangled. Thus, we consider an initial two-photon state given by  
\begin{equation}
|\Psi\rangle = [a|0_s0_i\rangle+b|1_s1_i\rangle]_{\mathbf{1}}\otimes|0_s0_i\rangle_{\mathbf{2}},
\end{equation}
where $|a|^2+|b|^2=1$ and $\{|0\rangle,|1\rangle\}$ is the logic basis for each qubit. The first step to obtain a HES is to couple the two DOFs. A coupling operation can be implemented through a single-photon two-qubit (SPTQ) \textsc{cnot} gate \cite{Fiorentino04} placed on one of the photon's arm (say signal). In this case the evolution of signal's DOF target will be conditioned to the state of its DOF control and the two-photon state becomes, after the \textsc{cnot} gate,  
\begin{equation}
|\Psi'\rangle = a|0_s0_i\rangle_{\mathbf{1}}|0_s\rangle_{\mathbf{2}}+b|1_s1_i\rangle_{\mathbf{1}}|1_s\rangle_{\mathbf{2}}.
\end{equation}
At this point the initial entanglement of DOF~$\mathbf{1}$ is diluted into this two-photon three-qubit Greenberger-Horne-Zeilinger (GHZ) state, as well as the entanglement between idler's DOF~$\mathbf{1}$ and signal's DOF~$\mathbf{2}$. To create hybrid entanglement we can either apply the \emph{irreversible eraser} [Fig.~\ref{fig:setup}(a)], which consists of a suitable measurement on signal's DOF~$\mathbf{1}$ [e.g., a projection onto $(|0_{s}\rangle+|1_{s}\rangle)_{\mathbf{1}}/\sqrt{2}$], or the \emph{reversible eraser} [Fig.~\ref{fig:setup}(b)], where a second SPTQ \textsc{cnot} gate is implemented, but now with the roles of the DOF control and target interchanged. After either one we get the hybrid entangled state:
\begin{equation}
|\Psi_\mathrm{HES}\rangle = a|{0_s}\rangle_{\mathbf{2}}|{0_i}\rangle_{\mathbf{1}}+b|{1_s}\rangle_{\mathbf{2}}|{1_i}\rangle_{\mathbf{1}}.
\end{equation}
Note that the irreversible eraser method creates HESs in a probabilistic manner and thereby it has the drawback of spending half of the photon pairs with the projection on the DOF control. On the other hand, in the reversible case it is possible to create HESs in a deterministic way with no loss of photons. 

	Given that the process of hybrid entanglement creation comprises the conversion of a DOF of one photon of the entangled pair into another, the HESs source will inherit all capabilities of the source whose DOF is the control in the protocol. Therefore, it is possible to create a tunable source of HESs if one has a tunable source of entanglement in the DOF control. For two-qubit states, tunable entanglement based on SPDC has already been demonstrated for polarization \cite{Kwiat99}, time-bin \cite{Thew02}, and transverse momentum \cite{Neves07} DOF. The same conclusion can be drawn, if one has an universal source which creates any type of two-qubit entangled state in the DOF control. Finally, even if the entanglement in the DOF control is not tunable, it is still possible to create pure nonmaximally HESs or maximally entangled ones from the initial state provided by the source, through a suitable erasure projection. Of course, this comes with the cost of reducing even more the ensemble size. We shall see these aspects in more detail along the discussion of our experiment.

\subsection{Our implementation}
\begin{figure}[t]
\centerline{\rotatebox{-90}{\includegraphics[width=0.325\textwidth]{{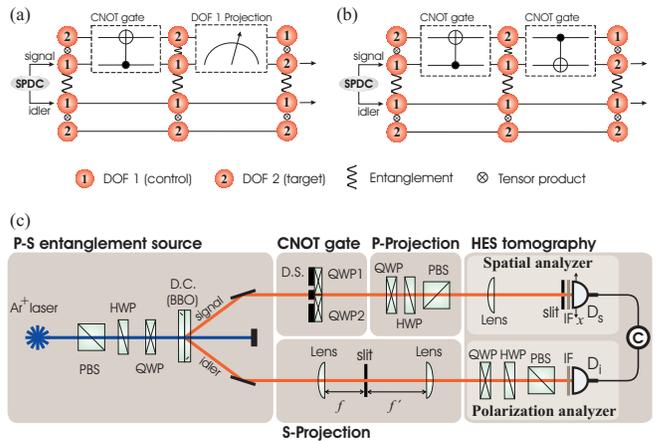}}}}
\caption{\label{fig:setup}(Color online) Schemes to create hybrid photonic entanglement using the (a) irreversible and (b) reversible disentanglement eraser phenomenon \cite{Garisto99}. (c) Outline of the experimental setup that implements the scheme (a) to create and characterize HESs. For details see text. P(S): polarization (spatial) DOF; PBS: polarizing beam splitter; HWP: half-wave plate; QWP: quarter-wave plate; DC: double type-I crystals; D.S.: double slit; IF: interference filter; D$_j$ ($j=s,i$) single-photon detectors; C: coincidence counter. } 
\end{figure}
\begin{SCfigure*}[0.55]
\centering\rotatebox{-90}{\includegraphics[width=0.4\textwidth]{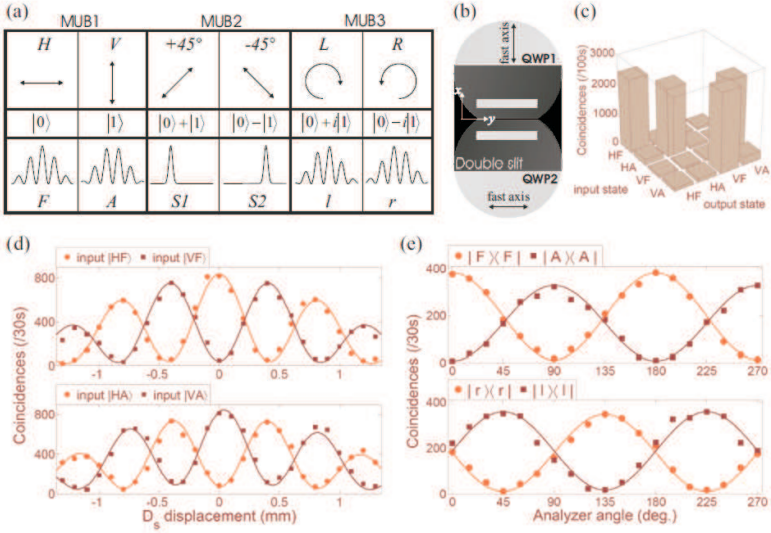}}
\hspace{0.8cm}
\caption{\label{fig:cnot}(Color online) (a) Comparison of polarization MUBs (first row) and their counterparts in spatial DOF (third row); the second row shows their (non-normalized) logical basis representation. (b) Double slit and the QWPs that implement a SPTQ \textsc{cnot} gate. (c)  Truth table with the measured amplitudes. (d)--(e) Measurements (symbols) and theoretical predictions (lines) of (d) the output spatial qubits for the input given in the legend and (e) the polarization and spatial DOF entanglement for the spatial projections given in the legend. The statistical errors of (d) and (e) are on the order of the symbol size in the figures.}
\end{SCfigure*}

	Our scheme to generate HESs is sketched in Fig.~\ref{fig:setup}(c) and employs photon pairs initially hyperentangled in both polarization and transverse momentum and position. As described above, this is not a requirement of the scheme and we use it here just for experimental convenience (see Sec.~\ref{sec_exp_a}). In this case we have just to perform an extra projection onto the logical basis of the target qubit (see discussion bellow), which would not be necessary starting with an non-hyperentangled state. We generate spatial-polarization hybrid entanglement from SPDC in a two-crystal geometry \cite{Kwiat99}. When one photon of the down-converted pair goes through a double slit and the transverse walk-off is negligible \cite{Osorio07} as in our setup, the following degenerate two-photon hyperentangled state is created \cite{Kwiat99,Neves07}:  
\begin{equation}
|\Psi\rangle=[a|H\rangle_s|H\rangle_i+b|V\rangle_s|V\rangle_i]\otimes[c|F\rangle_s|F\rangle_i+d|A\rangle_s|A\rangle_i],
\label{SOURCE}
\end{equation}
where $H$ ($V$) denotes horizontal (vertical) polarization and $F$ and $A$ denote orthogonal spatial modes defined by the double slit. The coefficients satisfy $|a|^2+|b|^2=1$ and $|c|^2+|d|^2=1$. Polarization entanglement is tuned by rotating the pump polarization through a half-wave plate (HWP); a quarter-wave plate (QWP) sets the phase \cite{Kwiat99}. Spatial entanglement can be tuned by manipulating the pump transverse profile \cite{Neves07}.  Figure~\ref{fig:cnot}(a) shows a comparison between the polarization mutually unbiased bases (MUBs) and their counterparts in the spatial DOF \cite{Lima08}, which will be useful for the discussions that follow. 

	The use of a hyperentanglement source with tunable entanglement in both DOF allows us to choose which one will be the control and the target in the protocol. For ease of manipulation, we chose here the polarization. On the other hand, for an irreversible eraser as we will implement, this might demand a projection on the DOF target, which will reduce the size of the ensemble for the HES preparation. A spatial filter in the idler arm performs this projection [Fig.~\ref{fig:setup}(c)].  

	The SPTQ \textsc{cnot} gate here is realized with a quarter-wave plate behind each slit of the double slit, with their fast axes orthogonally oriented as shown in Fig.~\ref{fig:cnot}(b). With polarization as the control qubit and spatial mode as the target qubit, the following conditional operations apply in the far field (focal or Fourier transform plane of a lens): 
\begin{eqnarray}
|H\rangle_s|F\rangle_s & \Rightarrow & |H\rangle_s|F\rangle_s, \nonumber\\
|H\rangle_s|A\rangle_s & \Rightarrow & |H\rangle_s|A\rangle_s,  \nonumber\\ 
|V\rangle_s|F\rangle_s & \Rightarrow & i|V\rangle_s|A\rangle_s , \nonumber\\ |V\rangle_s|A\rangle_s & \Rightarrow & i|V\rangle_s|F\rangle_s . \label{eq:cnot}
\end{eqnarray}
Therefore, these wave plates perform the same operation as an ideal \textsc{cnot} gate, up to a single-qubit phase shift. From Eq.~(\ref{SOURCE}) one can see that if the idler is projected onto $|F\rangle$, the two-photon state after the \textsc{cnot} gate becomes 
\begin{equation}
|\Psi_F\rangle = a|HF\rangle_s|H\rangle_i + ib|VA\rangle_s|V\rangle_i.  
\end{equation}
In order to create the hybrid entanglement, we measure the signal polarization. For a general polarization projector $|P\rangle\langle P|$, with $|P\rangle=\alpha|H\rangle+\beta|V\rangle$ ($|\alpha|^2+|\beta|^2=1$), we get the HES
\begin{equation} \label{hesgen}
|\Psi_F^{(P)}\rangle = \frac{1}{N}[a\alpha^*|F\rangle_s|H\rangle_i +i b\beta^*|A\rangle_s|V\rangle_i],
\end{equation} 
where $N=\sqrt{|a\alpha|^2+|b\beta|^2}$. By choosing $|P\rangle$ to be \emph{any} vector with $|\alpha|=|\beta|=1/\sqrt{2}$ [e.g., the MUBs 2 or 3 in Fig.~\ref{fig:cnot}(a)], the HES created has the same degree of entanglement of the initial polarization-entangled state. In this case, the tunable polarization-entanglement source will turn into a HESs source whose degree of entanglement can be continuously tuned by rotating the pump polarization. Obviously, there will be a cost of half of the photon pairs with this projection, as we discussed before. 

Following the conclusions drawn in Sec.~\ref{sec_theory_a}, we see that our source is able, in principle, of generating not only arbitrary two-qubit pure HESs but also tunable Werner states \cite{White01} and maximally entangled mixed states \cite{Peters04}, which have already been demonstrated for polarization-entanglement in the two-crystal source. Moreover, it is possible to realize a concentration-like process and filtering maximally HESs from initial nonmaximally polarization-entangled states (at the cost of reducing even more the ensemble size). From Eq.~(\ref{hesgen}) this is achieved when the polarization projector, $|P\rangle\langle P|$, has $|\alpha|=|b|$.

\section{Experiment}  \label{sec_exp}

\subsection{Experimental setup}   \label{sec_exp_a}
	Figure~\ref{fig:setup}(c) shows the experimental setup used to create and characterize photonic HESs. A 351.1~nm single-mode Ar$^+$-ion laser is used to pump, with 150~mW, two adjacent, orthogonally oriented, 0.5~mm thick $\beta$-barium borate (BBO) type-I crystals. The collimated  pump beam has a Gaussian  transverse profile with $\sim$2~mm diameter at the crystals. This profile guarantees the spatial entanglement of the two-photon state \cite{Neves07}. We could have focused the pump tightly into the crystal and get a separable spatial state. However, this would introduce impurity in the polarization-entangled state due to the walk-off effect \cite{Osorio07}.

Downconverted signal and idler photons travel through the setup at an angle of $3^{\circ}$ with the pump, and are spectrally filtered with a 10~nm interference filter (IF) centered at 702~nm, and detected by two single-photon detectors. The signal detector (D$_s$) is equipped with a 50~$\mu$m~$\times$~1.5~mm rectangular slit ($xy$), and mounted on a translation stage that displaces it in the $x$ direction; D$_i$ has a 1.2~mm circular aperture in front of it. Singles and coincidences are measured in a counter (C) with a resolving time of 5~ns. In the signal arm, a double slit aperture with a slit width of $80$~$\mu$m and center-to-center separation of $250$~$\mu$m is placed at 26~cm from the crystal. Behind each slit a QWP is fixed as shown in Fig.~\ref{fig:cnot}(b). After transmission, the photon pass through a polarization analyzer comprised of a QWP, HWP, and PBS, which is kept fixed for a suitable erasure projection. Then, it is collected by a lens system which combined with the transverse position of D$_s$, enables measurements of the spatial qubit in the MUBs shown in Fig.~\ref{fig:cnot}(a), either by imaging the double slit (MUB~2) or by  Fourier transforming it (MUBs~1 and 3) \cite{Lima08}. In the idler arm the spatial filter is comprised of a 100~$\mu$m~$\times$~3~mm slit placed at the focal planes of a $f=30$~cm focal length lens and a $f'=50$~cm focal length collimating lens. The slit is mounted on a translation stage that displaces it in the $x$ direction and set the spatial projection onto the basis $\{|F\rangle,|A\rangle\}$. After spatial filtering, a polarization analyzer (QWP, HWP, and PBS) allows polarization measurements in any basis.

\subsection{CNOT characterization}
Following Ref.~\cite{Fiorentino04} we characterize the \textsc{cnot} gate by showing its expected behavior and capacity of creating entanglement of the SPTQ state \cite{Comment2}. Figure~\ref{fig:cnot}(c) shows the truth table amplitudes where for each input state, the polarization analyzer and D$_s$ are  kept fixed to project onto each one of the four output states. Figure~\ref{fig:cnot}(d)  shows the output spatial states, measured by displacing D$_s$ in the $x$ direction at the Fourier transform plane, for the input states given in the legend. In both figures the predicted behavior is clearly seen. Finally, we show in Fig.~\ref{fig:cnot}(e) the capacity of the \textsc{cnot} gate to entangle the spatial and polarization DOF of the signal photon. A HWP is inserted before the double slit and prepares the input state $[|H\rangle_s+|V\rangle_s]|F\rangle_s/\sqrt{2}$. After transmission we get $[|HF\rangle_s+i|VA\rangle_s]/\sqrt{2}$. For each spatial projection shown in the legends, a linear polarization analyzer is rotated and coincidences recorded. The results shown in the second panel are a signature that a $\pi/2$ phase shift is introduced by our \textsc{cnot} gate.

\subsection{Hybrid entangled states characterization}

\begin{figure*}[t]
\centerline{\rotatebox{-90}{\includegraphics[width=0.67\textwidth]{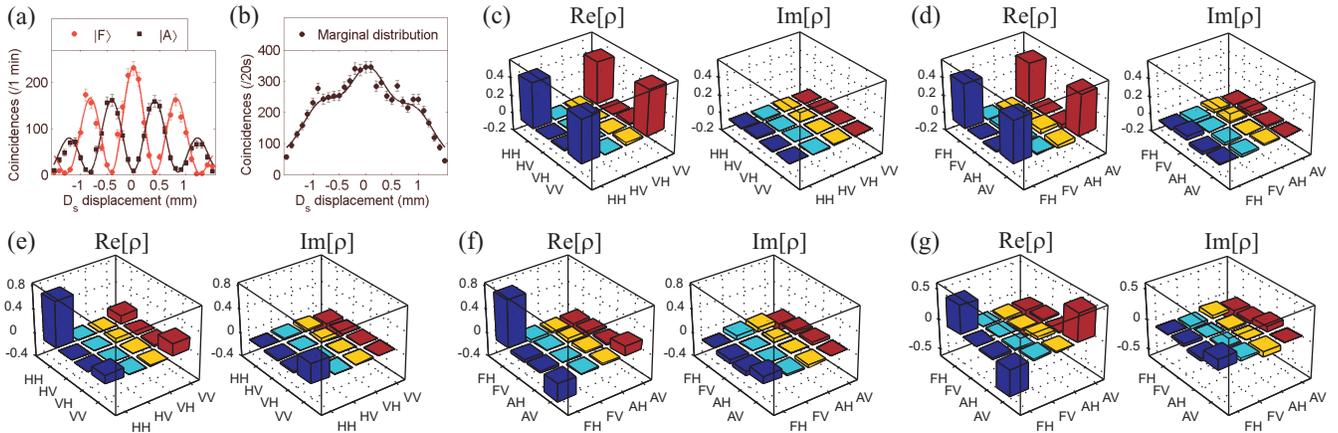}}}
\vspace{-6.3cm}
\caption{\label{fig:hesmes}(Color online). (a)--(b) Measurements (symbols) and theoretical predictions (lines) of the Schmidt modes (a) and marginal distribution (b) of the spatial part of Eq.~(\ref{SOURCE}). (c)--(g) Density matrix plots obtained by maximum likelihood estimation from the experimental data. (c) Initial maximally  polarization-entangled state and  (d) the corresponding HES with the erasure polarization projection onto $L$. (e) Initial partially polarization-entangled state, (f) the corresponding HES with the erasure projection onto  $45^{\circ}$, and (g) the filtered maximally HES with the erasure projection onto $62.5^{\circ}$.}
\end{figure*}
	For the pump profile we are using, the spatial part of the two-photon state [Eq.~(\ref{SOURCE})] is highly entangled. To set the action of the spatial filter we must characterize this state in order to reduce the losses due to this filtering. With a double slit without the QWPs behind it, we traced over the polarization and measured the spatial Schmidt modes [Fig.~\ref{fig:hesmes}(a)] and the marginal distribution [Fig.~\ref{fig:hesmes}(b)]. The former are measured by projecting the idler onto $|F\rangle$ and $|A\rangle$ and the latter by tracing over idler spatial DOF and scanning D$_s$ in both cases.  Accordingly with \cite{Neves07}, the concurrence is evaluated from the number of coincidences in each pattern in the first case and from the visibility of the interference in the second case. Therefore, we obtained a nearly maximal degree of spatial entanglement, namely,  $0.99\pm 0.12$ and $0.99\pm 0.01$, respectively. We then set the spatial filter in the idler arm to select the $|F\rangle$ state, for all results that follow. 

	The initial polarization-entangled states are prepared without the QWPs behind the double slit and characterized, by quantum tomography, with D$_s$ fixed in $x=0$ at the focal plane of the lens. The HESs are characterized  by combining the methods of polarization \cite{James01} and spatial qubits \cite{Lima08} tomography. Figures~\ref{fig:hesmes}(c) and \ref{fig:hesmes}(d) show the density matrix plots of the initial maximally polarization-entangled state and the corresponding HES, respectively. The former has a fidelity of $\mathcal{F}_\mathrm{pol}=96.9\pm 2.5\%$ with the Bell state $[|H\rangle_s|H\rangle_i+|V\rangle_s|V\rangle_i]/\sqrt{2}$, where $\mathcal{F}\equiv \langle\Psi|\rho|\Psi\rangle$ ($\rho$ measured and $|\Psi\rangle$ expected state). The latter, which is obtained through an erasure projection onto $L$, has $\mathcal{F}_\mathrm{HES}=92.7\pm 3.2\%$ with the Bell HES $[|F\rangle_s|H\rangle_i+|A\rangle_s|V\rangle_i]/\sqrt{2}$. Data in Figs.~\ref{fig:hesmes}(c) and \ref{fig:hesmes}(d) were collected for 300 and 600~s per projection with count rates of 4.2 and 1.8~s$^{-1}$, respectively. 

	Next we started with a nonmaximally polarization-entangled state. It is shown in Fig.~\ref{fig:hesmes}(e) and has a fidelity of $\mathcal{F}_\mathrm{pol}=96.5\pm 2.8\%$ with the state $0.89|H\rangle_s|H\rangle_i+0.46e^{i\phi}|V\rangle_s|V\rangle_i$ ($\phi=0.37\pi$). With an erasure projection onto $45^{\circ}$ we get the corresponding partially HES shown in Fig.~\ref{fig:hesmes}(f) which has $\mathcal{F}_\mathrm{HES}=93.8\pm 2.8\%$ with $0.89|F\rangle_s|H\rangle_i+0.46e^{i\phi}|A\rangle_s|V\rangle_i$ ($\phi=0.87\pi$). This shows that our HESs source is tunable. We also show that a maximally HES can be filtered from this initial partially polarization-entangled state. Here, this is achieved through an erasure projection onto $62.5^{\circ}$ ($\alpha=0.46$). The resultant state is shown in Fig.~\ref{fig:hesmes}(g) and has $\mathcal{F}_\mathrm{HES}=92.8\pm 3.1\%$ with $[|F\rangle_s|H\rangle_i+e^{i\phi}|A\rangle_s|V\rangle_i]/\sqrt{2}$ ($\phi=0.87\pi$). Data in Figs.~\ref{fig:hesmes}(e)--\ref{fig:hesmes}(g) were collected for 300, 600, and 1000~s per projection with count rates of 4.2, 1.8, and 0.98~s$^{-1}$, respectively. 

	The reduction in the fidelity for all the HESs generated is mainly due to the nonperfect alignment of the QWPs behind the double slit which leads to an imperfect coupling between polarization and spatial DOF.

\section{Applications} \label{sec_app}
	Since entangled states play a key role in many applications of QC as well as in fundamental tests of quantum mechanic, any source of photonic entanglement is useful in principle, no matter in which DOF the quantum information is encoded. By this we are advocating the usefulness of HESs like any other single DOF entanglement. In the latter case, the different encodings present practical advantages and disadvantages when compared to each other, which can be attested by the extensive research of sources for polarization, time-bin or spatially entangled photon pairs. In much the same way, hybrid photonic entanglement, when compared with entanglement in a single DOF, brings some advantages, as we will discuss below, and possesses its own drawbacks, as for instance the loss of photons when the reversible eraser is employed. Next, we give examples where a HESs source could find applications, beyond the fundamental tests previously reported \cite{Zukowski91,Ma09,Neves09}.

\subsection{Engineering qubit-qudit entanglement}
	The entanglement between a qubit and a qudit using photons is largely unexplored mainly because it is not trivial to create. The only report in the literature, has demonstrated a scheme where a measurement-induced nonlinearity is employed on three-photon polarization states to generate entangled qubit-qutrit systems ($D=3$) only \cite{Lanyon08}. Following a recent interest for developing quantum gates \cite{Daboul03} and determining entanglement measures \cite{Chen05} and dynamics \cite{Ann07,Li09} in such qubit-qudit systems, it would be worthy to find another method to  create them optically. When compared with \cite{Lanyon08}, such method should satisfy the following requirements: (i) it is easier with respect to the operations employed, (ii) it is less costly with respect to the number of photons used, and (iii) it is more flexible with respect to the attainable dimension of the qudit. Here, we show that HESs satisfy these requirements, and our setup in particular can be easily extended to create a tunable source of qubit-qudit entangled states with $D\gg 3$.

	We restrict ourselves to qudits with even dimension, i.e., $D=2n$ ($n\geq 2$), because in the odd case it would not be possible the creation of maximal entanglement with the QWPs only. If one replaces the double slit in Fig.~\ref{fig:setup}(c) by a $D$-slit aperture, a spatial qudit is created \cite{Neves05}. By fixing each of the orthogonally oriented QWPs behind $n$ slits, we have, generalizing (\ref{eq:cnot}), 
\begin{eqnarray}
|H\rangle_s|F_j\rangle_s & \Rightarrow & |H\rangle_s|F_j\rangle_s, \nonumber\\
|V\rangle_s|F_j\rangle_s & \Rightarrow & i|V\rangle_s|F_{j\oplus n}\rangle_s,
\end{eqnarray}
where $\oplus$ denotes addition modulo $D$ and $\{F_j\}$ ($j=0,\ldots,D-1$), $D$ orthogonal spatial modes of the qudit at the Fourier transform plane, defined by the equally weighted superposition of the $D$ available paths (slits $\{|l\rangle\}$) as $\sum_{l=0}^{D-1}e^{i\phi_{l}^{(j)}}|l\rangle/\sqrt{D}$, with $\phi_{l}^{(j)}=0$ or $\pi$ \cite{Neves07,Neves05}. Assuming an initial two-photon state $[a|H\rangle_s|H\rangle_i+b|V\rangle_s|V\rangle_i]\otimes|F_j\rangle_s|F_j\rangle_i,$ and an erasure polarization projection onto $L$ after the \textsc{cnot} gate, we get the following two-photon qubit-qudit HES:
\begin{equation}  \label{bitdit}
|\Psi_{F_j}^{(L)}\rangle = a|F_j\rangle_s|H\rangle_i + b|F_{j\oplus n}\rangle_s|V\rangle_i.
\end{equation}
Therefore, the scheme proposed here employs only \emph{two} photons, simple local operation, it is easily extended to higher dimensions (up to $D=8$ at least \cite{Neves05}) and the degree of entanglement can be easily tuned from zero to maximal. It is clearly advantageous compared with \cite{Lanyon08}. 

	The generation of qubit-qudit states like Eq.~(\ref{bitdit}) is useful for determining experimentally the dynamics of entanglement for any state of such systems when the \emph{qubit} goes through an arbitrary noisy channel \cite{Cirone03}. It has been demonstrated in \cite{Li09} (generalizing the result of \cite{Konrad08} for two qubits) that the evolution of entanglement for an arbitrary qubit-qudit initial state is determined only by the action of the channel on the maximally entangled state.

\subsection{Hybrid optical quantum network}
	An optical quantum network employs free-space or optical fiber channels for transmitting quantum information. Either one is decided mainly by the robustness of the photonic DOF along the channel. Therefore, a HESs source can be advantageous as it enables a more flexible network with each photon being transmitted through the more suited channel. For instance, polarization qubits at optical wavelengths suffer decoherence in transmissions along optical fibers, but are faithfully transmitted through atmosphere \cite{Resch05} due the non-birefringent nature of the latter.  On the other hand the atmospheric turbulence may affect the spatial DOF \cite{Paterson05}, and despite of its multimode nature, it could be coupled into a bundle of single-mode fibers \cite{Rossi09} and analyzed through a multiport beam splitter \cite{Zukowski97} after transmission. In particular, for fiber transmission the time-bin is more suited \cite{Thew02} as it is not affected by the thermal or mechanical fiber instabilities (which can affect the spatial DOF). Since the scheme for HESs generation we described here can be applied to any DOF, it could also be used.

\section{Conclusion} \label{sec_conc}

	In summary, we have shown that the quantum disentanglement eraser is a general method to produce hybrid photonic entanglement which can be realized with any degree of freedom. 
To demonstrate this we created and characterized a SPDC source of tunable spatial-polarization HESs. Moreover, the applicability and usefulness of HESs have been pushed beyond the fundamental tests of quantum theory, previously reported. We showed that HESs allow for: (i) an easier, more flexible and less costly generation of qubit-qudit entangled photons, in comparison with a previous scheme \cite{Lanyon08} and (ii) a more flexible optical quantum network with free-space and optical fiber channels.

	Another interesting aspect of hybrid photonic systems, introduced here, is that they may be the natural route for the optical implementation of quantum gates between two or more qudits of different dimensions \cite{Daboul03}. Finally, we would like to emphasize that in the same way we proposed the extension of our setup with a $D$-slit array to create qubit-qudit entanglement, there is no restriction, in principle, to extend it to a continuous spatial domain and create photonic qubit-continuous variable entanglement and hence, establish a connection between quantum information protocols with discrete and continuous variables. This type of entanglement is  observed, for instance, in atoms in high-$Q$ cavities \cite{Brune96} and trapped-ion systems \cite{Monroe96}.

\textit{Note added}.-- The experimental implementation \cite{Ma09} of the Bell inequality with the HES proposed in \cite{Zukowski91} has recently become available.

\begin{acknowledgments}
We would like to thank C. H. Monken for lending us the sanded quarter-wave plates used in the double slit. This work was supported by Milenio (Grant No. ICM P06-067F), PBCT (Grant No. PDA-25), and FONDECYT (Grant No. 11085057). G.L. and A.D. acknowledge the support by FONDECYT (Grants No. 11085055 and No. 1061046, respectively).
\end{acknowledgments}


\begin{thebibliography}{xxxx}

	\bibitem{Bouwmeester97} D. Bouwmeester, J.-W. Pan, K. Mattle, M. Eibl, H. Weinfurter, and A. Zeilinger, Nature (London) \textbf{390}, 575 (1997).

	\bibitem{Jennewein00} T. Jennewein, C. Simon, G. Weihs, H. Weinfurter, and A. Zeilinger, \prl \textbf{84}, 4729 (2000).

	\bibitem{Weihs98} G. Weihs, T. Jennewein, C. Simon, H. Weinfurter, and A. Zeilinger, \prl \textbf{81}, 5039 (1998).

	\bibitem{Kwiat99} P. G. Kwiat, E. Waks, A. G. White, I. Appelbaum, and P. H. Eberhard, \pra \textbf{60}, R773 (1999). 

	\bibitem{Lamas01} A. Lamas-Linares, J. C. Howell, and D. Bouwmeester, Nature (London) \textbf{412}, 887 (2001).

	\bibitem{Thew02}  R. T. Thew, S. Tanzilli, W. Tittel, H. Zbinden, and N. Gisin, \pra \textbf{66}, 062304 (2002).

	\bibitem{Riedmatten04} H. de Riedmatten, I. Marcikic, V. Scarani, W. Tittel, H. Zbinden, and N. Gisin, \pra \textbf{69}, 050304(R) (2004). 

	\bibitem{Neves07} L. Neves, G. Lima, E. J. S. Fonseca, L. Davidovich, and S. P\'adua, \pra \textbf{76}, 032314 (2007).  

	\bibitem{Neves05} L. Neves, S. P\'adua, and C. Saavedra, \pra \textbf{69}, 042305 (2004); L. Neves, G. Lima, J. G. Aguirre G\'omez, C. H. Monken, C. Saavedra, and S. P\'adua, \prl \textbf{94}, 100501 (2005). 

	\bibitem{Rossi09} A. Rossi, G. Vallone, A. Chiuri, F. De Martini, and P. Mataloni, \prl \textbf{102}, 153902 (2009).

	\bibitem{Wieczorek08} W. Wieczorek, C. Schmid, N. Kiesel, R. Pohlner,
O. G\"uhne, and H. Weinfurter, \prl \textbf{101}, 010503 (2008).

	\bibitem{Barreiro05} J. T. Barreiro, N. K. Langford, N. A. Peters, and P. G. Kwiat, \prl \textbf{95}, 260501 (2005).

	\bibitem{Zukowski91} M. \.{Z}ukowski and A. Zeilinger, Phys. Lett. A \textbf{155}, 69 (1991).

	\bibitem{Ma09} X.-s. Ma, A. Qarry, J. Kofler, T. Jennewein, and A. Zeilinger, \pra \textbf{79}, 042101 (2009).

	\bibitem{Neves09} L. Neves, G. Lima, J. Aguirre, F. A. Torres-Ruiz, C. Saavedra, and A. Delgado, New J. Phys. \textbf{11}, 073035 (2009). 

	\bibitem{Garisto99} R. Garisto and L. Hardy, \pra \textbf{60}, 827 (1999).

	\bibitem{Fiorentino04} M. Fiorentino and F. N. C. Wong, \prl \textbf{93}, 070502 (2004).

	\bibitem{Osorio07} C. I. Osorio, G. Molina-Terriza, B. G. Font, and J. P. Torres, Opt. Express \textbf{15}, 14636 (2007). 

	\bibitem{Lima08} G. Lima, F. A. Torres-Ruiz, L. Neves, A. Delgado, C. Saavedra, and S. P\'adua, J. Phys. B \textbf{41}, 185501 (2008); G. Taguchi, T. Dougakiuchi, N. Yoshimoto, K. Kasai, M. Iinuma, H. F. Hofmann, and Y. Kadoya, \pra \textbf{78}, 012307 (2008). 

	\bibitem{White01} A. G. White, D. F. V. James, W. J. Munro, and P. G. Kwiat, \pra \textbf{65}, 012301 (2001).

	\bibitem{Peters04} N. A. Peters, J. B. Altepeter, D. A. Branning, E. R. Jeffrey, T.-C. Wei, and P. G. Kwiat, \prl \textbf{92}, 133601 (2004).

	\bibitem{Comment2} For the \textsc{cnot} characterization, we focused the pump into the BBO crystals (waist size $\sim$65~$\mu$m)  and prepared the input spatial target states by tilting the double slit around the $y$ axis. The transverse walk-off is not an issue here since the input polarization states are separable \cite{Osorio07}.

	\bibitem{James01} D. F. V. James, P. G. Kwiat, W. J. Munro, and A. G. White, \pra  \textbf{64}, 052312 (2001).

	\bibitem{Lanyon08} B. P. Lanyon, T. J. Weinhold, N. K. Langford, J. L. O'Brien, K. J. Resch, A. Gilchrist, and A. G. White, \prl \textbf{100}, 060504 (2008).

	\bibitem{Daboul03} J. Daboul, X. Wang,  and B. C. Sanders, J. Phys. A \textbf{36}, 2525 (2003).

	\bibitem{Chen05} K. Chen, S. Albeverio, and S.-M. Fei, \prl \textbf{95}, 210501 (2005).

	\bibitem{Ann07} K. Ann and G. Jaeger, Phys. Lett. A \textbf{372}, 579 (2008).

	\bibitem{Li09} Z.-G. Li, S.-M. Fei, Z. D. Wang, and W. M. Liu, \pra \textbf{79}, 024303 (2009).

	\bibitem{Cirone03} M. A. Cirone, A. Delgado, D. G. Fischer, M. Freyberger, H. Mack, and M. Mussinger, Quantum Inf. Process. \textbf{1}, 303 (2002).

	\bibitem{Konrad08} T. Konrad, F. De Melo, M. Tiersch, C. Kasztelan, A. Arag\~ao,
and A. Buchleitner, Nature Phys. \textbf{4}, 99 (2008).

	\bibitem{Resch05} K. J. Resch \emph{et al.}, Opt. Express \textbf{13}, 202 (2005). 

	\bibitem{Paterson05} C. Paterson, \prl \textbf{94}, 153901 (2005).

	\bibitem{Zukowski97} M. \.{Z}ukowski, A. Zeilinger, and M. A. Horne, \pra \textbf{55}, 2564 (1997).

	\bibitem{Brune96} M. Brune, E. Hagley, J. Dreyer, X. Ma\^itre, A. Maali, C. Wunderlich, J. M. Raimond, and S. Haroche, \prl \textbf{77}, 4887 (1996).

	\bibitem{Monroe96} C. Monroe, D. M. Meekhof, B. E. King, and D. J. Wineland, Science \textbf{272}, 1131 (1996).

\end{thebibliography}
\end{document}